\documentclass[prd,10pt,aps,twocolumn,amsfonts,showpacs,superscriptaddress,nofootinbib]{revtex4-2}
\usepackage[ascii]{inputenc}
\usepackage{amsmath,amssymb,amsfonts}
\usepackage{graphicx}
\usepackage[caption=false]{subfig}
\usepackage{xcolor}
\usepackage{ulem}
\usepackage[bookmarks=false,colorlinks=true,urlcolor=violet,linkcolor=blue,citecolor=red,hyperindex=true,linktocpage=true]{hyperref}

\newcommand{\mmu}{\mu}
\newcommand{\mpi}{\pi}

\graphicspath{{figures/}}

\begin{document}

\title{Coherent Terahertz Radiation from a Nonlinear Oscillator of Viscous Electrons}

\author{Christian B.~Mendl}
\email{christian.mendl@tum.de}
\affiliation{Technische Universit\"at Dresden, Institute of Scientific Computing, Zellescher Weg 12-14, 01069 Dresden, Germany}
\affiliation{Technische Universit\"at M\"unchen, Department of Informatics and Institute for Advanced Study, Boltzmannstra{\ss}e 3, 85748 Garching, Germany}

\author{Marco Polini}
\email{marco.polini@iit.it}
\affiliation{Dipartimento di Fisica dell'Universit\`a di Pisa, Largo Bruno Pontecorvo 3, I-56127 Pisa, Italy}
\affiliation{School of Physics \& Astronomy, University of Manchester, Oxford Road, Manchester M13 9PL, United Kingdom}
\affiliation{Istituto Italiano di Tecnologia, Graphene Labs, Via Morego 30, I-16163 Genova, Italy}

\author{Andrew Lucas}
\email{andrew.j.lucas@colorado.edu}
\affiliation{Department of Physics, Stanford University, Stanford CA 94305, USA}
\affiliation{Department of Physics, University of Colorado, Boulder CO 80309, USA}

\date{\today}

\begin{abstract}
Compressible electron flow through a narrow cavity is theoretically unstable, and the oscillations occurring during the instability have been proposed as a method of generating Terahertz radiation. We numerically demonstrate that the endpoint of this instability is a nonlinear hydrodynamic oscillator, consisting of an alternating shock wave and rarefaction-like relaxation flowing back and forth in the device. This qualitative physics is robust to cavity inhomogeneity and changes in the equation of state of the fluid. We discuss the frequency and amplitude dependence of the emitted radiation on physical parameters (viscosity, momentum relaxation rate, and bias current) beyond linear response theory, providing clear predictions for future experiments.
\end{abstract}

\maketitle

Generating and detecting Terahertz (THz) radiation (photon frequency in the range $\sim 100~{\rm GHz}$ -- $30~{\rm THz}$) are challenging tasks~\cite{Dhillon_jphysD_2017}. The reason is that THz radiation lies in the gap between the realms of photonics (on the high-frequency side) and of electronics (on the low-frequency range). Tremendous progress in the generation of THz radiation has been achieved in the last two decades~\cite{Dhillon_jphysD_2017}. A number of THz sources have indeed been realized, ranging from THz quantum cascade lasers to ${\rm LiNbO}_3$-based optical rectifiers, nonlinear organic crystals, ionized plasmas, difference frequency mixers of optical parametric amplifiers, and accelerators. We refer the reader to Ref.~\cite{Dhillon_jphysD_2017} for a recent review. 

THz sources that are compact, powerful, tunable, and operating at room temperature are, however, still in great demand for many industrial applications as well as fundamental science, including studying excitations in solids and a plethora of non-equilibrium phenomena. To this end, active research \cite{Koppens,wang_nano_2018} targets the use of two-dimensional (2D) materials, such as graphene, and their van der Waals heterostructures~\cite{Geim2013,Mounet2018}. The main reasons are the small footprint of these materials, the broadband nature and gate tunability of graphene's optical properties \cite{Koppens}, and its exceptionally high electronic quality, unmatched by any other material at room temperature, when encapsulated in atomically flat insulators such as hexagonal Boron Nitride (hBN) \cite{mayorov_nanolett_2012,wang_science_2013}.

A longstanding proposal for the generation of coherent THz radiation comes from Dyakonov and Shur \cite{DS}, who noted that a 2D electron system in a hydrodynamic regime exhibits a compressible instability in a cavity with peculiar (but feasible) boundary conditions, which we describe below. In a separate work \cite{DS2}, Dyakonov and Shur proposed also to use hydrodynamic electron fluids to detect THz radiation (for recent work on the topic see Ref.~\cite{Svintsov}), which is converted to a dc electrical signal thanks to hydrodynamic nonlinearities. 

Despite many experimental attempts, Dyakonov-Shur (DS) {\it generation} of THz radiation has not been reported. On the contrary, downconversion of THz radiation to a dc signal due to a variety of nonlinearities has been measured \cite{tauk,giliberti,Vitiello,Vicarelli,Cai,Pablo,Bandurin_apl_2018}, even as resonantly enhanced by plasmons in high-quality hBN-encapsulated graphene devices \cite{bandurinDS}.

As electrons only behave hydrodynamically in ultra-pure crystals in materials with relatively small Fermi surfaces and strong electron-electron scattering \cite{gurzhi, andreev, tomadin, torre, levitov, lucas3, guo, scaffidi}, one possible obstruction to DS generation of THz radiation has been the challenge of observing the hydrodynamic regime in electron fluids, which has largely been impossible until recent discoveries in graphene \cite{bandurin, crossno, levitov1703, bandurin18, ku, sulpizio} and GaAs \cite{molenkamp, bakarov, braem}: see Ref.~\cite{lucasreview2} for a review.

Given the many experimental efforts to find such instability, and theoretical work suggesting its robustness in the presence of dissipation \cite{DS}, long-range Coulomb interactions \cite{DS05}, and even in the absence of strong electron-electron interactions \cite{DSkin, mendl}, we find it quite surprising that the effect has proven so challenging to find. The purpose of this work is to answer a simple question: is the instability not observable in experiments because it terminates in an exotic non-equilibrium ``state"? Does its endpoint differ in a significant enough way from linear response theory that it would not be detected in existing experimental setups?

We address this question by numerically solving the dissipative hydrodynamic equations for electron fluids in the appropriate geometry. Our principal finding is that the endpoint of instability {\it is a coherent, nonlinear oscillator}, consisting of alternating shock and rarefaction waves bouncing between the ends of the cavity. We analyze the properties of this nonlinear oscillator, including the amplitude and frequency of the emitted radiation, each of which depend on all physical parameters in the problem (viscosity, momentum relaxation rate, etc.).  Previous literature in the ideal hydrodynamic (dissipationless) limit has also reached similar conclusions \cite{cheremisin, kangli, cosme}. Our results suggest that instability {\it should be detectable} in present day experiments, and provide further predictions for experiments well beyond the linear stability analysis of Ref.~\cite{DS2}.

\begin{figure}[t]
\centering
\includegraphics[width=\columnwidth]{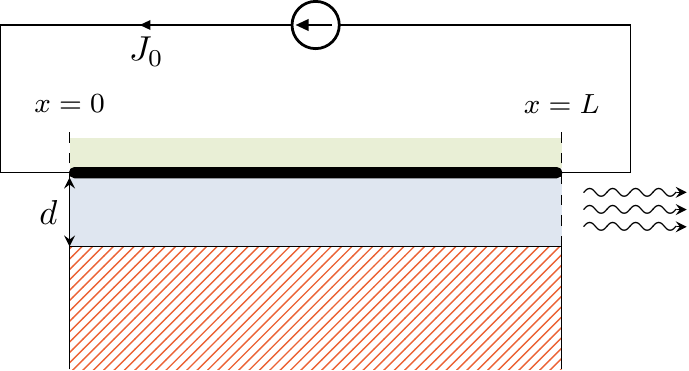}
\caption{A graphene layer (black) is electrically separated from a metal gate (red hatched); dielectrics such as hBN are usually used to fill the metal-graphene gap providing electrical isolation and also placed above graphene to screen it from impurities. The metal/graphene hybrid forms a capacitor whose charging and uncharging generates power in the THz range. The oscillatory dynamics originates from an instability caused by the presence of bias current in the graphene.}
\label{fig:setup}
\end{figure}

\medskip

We first describe the basic setup of \cite{DS} for generating THz radiation, depicted in Fig.~\ref{fig:setup}. A 2D material, such as graphene, is placed at a distance $d$ from a metallic gate. Dielectric media such as hBN are placed between graphene and the metal gate to provide electrical isolation and above graphene to screen it from contamination. The length of the graphene is $L$, and the width of the device is $w$. The graphene/gate act as an effective capacitor. As we will see, the oscillatory dynamics of the electron fluid charges and uncharges this capacitor, and the energy stored in the capacitor can be radiated away through an antenna.

Recent experiments \cite{bandurin, crossno, levitov1703, bandurin18, ku, sulpizio} demonstrate that the electrons in high quality graphene are well described for temperatures $70~{\rm K} \lesssim T \lesssim 300~{\rm K}$ by a hydrodynamic theory. We denote the electron number density $n$, thermodynamic pressure $P(n)$, shear viscosity $\eta$, momentum relaxation rate $\gamma$, and external electric field $E$. The charge of the electron is $-e$, and its effective mass is $m$. Assuming equilibrium charge density $n_0$ and fluid element velocity $v_0$, Newton's Law implies the existence of a steady state with dc current flow through the device, with electron number current (also called electron flux) \begin{equation}
J_0 = n_0 v_0 = \frac{-e n_0 E}{\gamma m}.
\end{equation}

In the laboratory, it is not possible to construct a flow with perfectly homogeneous density $n(x,t)=n_0$ and number current $J(x,t) \equiv n(x,t)v(x,t) = J_0$, where we have introduced the fluid element velocity $v(x,t)$. We model the dynamics of long wavelength perturbations to this steady state using nonlinear hydrodynamics:
\begin{subequations}\label{eq:hydroeq}\begin{align}
\partial_t n + \partial_x J &= 0, \\
\partial_t J + \partial_x \left(\frac{J^2}{n} + \frac{P(n)}{m} - \frac{\eta}{m} \partial_x \frac{J}{n}\right) &= \gamma (n v_0 - J).
\end{align}\end{subequations}
The first equation is the continuity equation, while the second one is the Navier-Stokes (momentum balance) equation. The one-dimensional modelling is justified by the translational invariance of the DS boundary conditions introduced below along the $y$ direction orthogonal to $x$ \cite{Principi}. In the main text, we assume the simplified relation
\begin{equation}
P(n) = mv_{\mathrm{s}}^2 n,\label{eq:pressure}
\end{equation} 
where $v_{\mathrm{s}}$ is the speed of sound in the fluid and taken to be independent of $n$. We also define the three relevant dimensionless quantities $\tilde{v}_0 = v_0/v_{\mathrm{s}}$, $\tilde\eta = \eta/(mn_0 v_{\mathrm{s}}L) \equiv \nu/(v_{\mathrm{s}} L)$---where we have introduced the kinematic viscosity $\nu=\eta/(n_0 m)$---and $\tilde\gamma = \gamma L/v_{\mathrm{s}}$. Eqs.~\eqref{eq:hydroeq} and \eqref{eq:pressure} form a minimal model for our hydrodynamic instability. The qualitative conclusions of our study are unchanged by more realistic equations of state $P(n)$, which we derive and discuss in the Supplementary Information (SI).

We study \eqref{eq:hydroeq} in the spatial domain $0 \le x \le L$, subject to the DS boundary conditions
\begin{subequations}\begin{align}
n(0) &= n_0, \\
J(L) &= n_0 v_0, \\
\partial_x J(0) &= 0.
\end{align}\end{subequations}
The quasinormal modes near equilibrium have frequencies 
\begin{equation}
\omega_n = \frac{\mpi v_{\mathrm{s}}n}{2L} + \mathrm{i}\left(\frac{v_0}{L} - \frac{\gamma}{2} - \frac{\mpi^2n^2}{8L^2} \nu\right) + \cdots, \label{eq:LRomega}
\end{equation}
where $\cdots$ denotes terms quadratic in $\gamma$, $\eta$, and $v_0$, and $n=1,3,5,\ldots$ is an odd integer. 
Clearly, whenever $\tilde v_0 > \frac{1}{2}\tilde\gamma + \frac{\pi^2}{8}\tilde\eta$ there is an instability. The main result of this paper is the numerical determination of the endpoint of this instability.  Since in a typical high-quality device, $v_{\mathrm{s}}\sim 10^6$ m/s and $L\sim 1$ $\mmu$m, the frequency $\omega_1 \sim 10^{12} \; \mathrm{s}^{-1}$ is in the THz range.

\begin{figure}[t]
\centering
\includegraphics[width=3.3in]{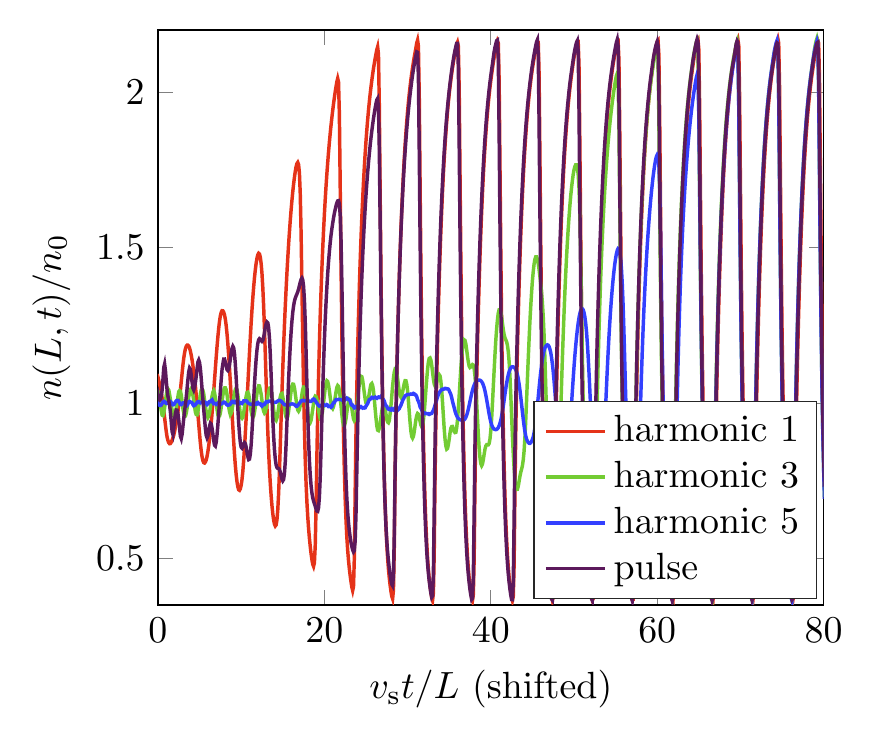}
\caption{The endpoint of the instability is universal and robust to initial conditions. Parameters are $\tilde v_0=0.14$, $\tilde\eta = 0.01$, and $\tilde\gamma=0.04$; for this choice, \eqref{eq:LRomega} predicts that the $n=3$ mode is just above the instability threshold. $t$ has been offset by a constant for each simulation so that oscillations appear in sync as $t\rightarrow \infty$. With a finite amplitude, sourcing the $n=5$ harmonic (formally stable) nevertheless ultimately leads to instability.\label{fig:universal}}
\end{figure}

We solved \eqref{eq:hydroeq} numerically using finite volume methods combined with Strang splitting (see SI for details). We choose parameters $v_0$, $\eta$, and $\gamma$ such that $\max(\mathrm{Im}(\omega_n))>0$ in order to observe the hydrodynamic instability. The endpoint of this instability is a coherent, nonlinear oscillator, which we numerically observe to be {\it universal}---neither the amplitude, frequency nor any other features of the nonlinear oscillations depend on the initial conditions: see Fig.~\ref{fig:universal}. This statement remains true even when $v_0$ is large. In the latter case, Eq.~\eqref{eq:LRomega} predicts that there are multiple normal modes which are unstable. Fig.~\ref{fig:universal} shows the dynamics of $n(L,t)$ for multiple different initial conditions at such high $v_0$. Generic initial conditions tend towards the same oscillatory attractor solution.

\begin{figure*}[t]
\centering
\includegraphics[width=0.95\textwidth]{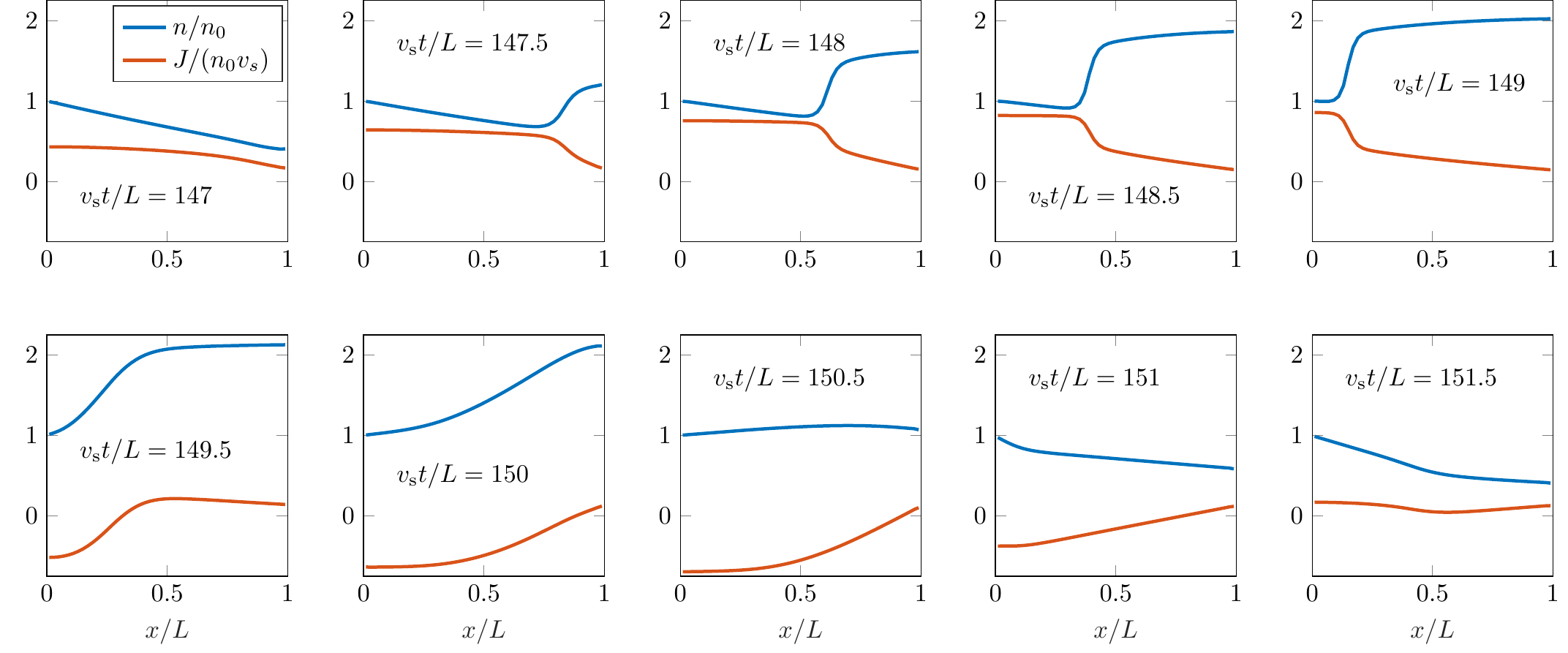}
\caption{The dynamics of the number density $n(x,t)$ and current $J(x,t)$ throughout a period of the nonlinear hydrodynamic oscillations at the endpoint of the instability. Parameters used are identical to Fig.~\ref{fig:universal}.}
\label{fig:snapshots}
\end{figure*}

Our simulations demonstrate unambiguously that the endpoint of the instability is a coherent, nonlinear oscillator. Within numerical resolution, there is a period $T$ such that as $t\rightarrow \infty$, $n(x,t)=n(x,t+T)$ and $J(x,t)=J(x,t+T)$. Ten snapshots of the local density and current in the cavity at different points in time are shown in Fig.~\ref{fig:snapshots}. We interpret these plots as depicting a left-moving shock wave (with wavefront smoothed by viscous effects), and a right-moving ``rarefaction wave".   Since $n(x=0)$ is fixed, charge must accumulate on the right as the shock wave propagates to the left (note $J(x=0)>J(x=L)$ at this time).  As this is an oscillator, at later times $J(x=0)<J(x=L)$ must arise, to drain the system of the built-up charge.  This occurs when the rarefaction wave moves to the right and causes a depletion of charge near $x=L$.  Similar physics is observed for more realistic equations of state, and for models of inhomogeneous cavities (see SI).  

\medskip

The remainder of this paper focuses on the application of this oscillatory attractor to THz radiation generation. In what follows, we propose future experiments in ultra-clean hBN/$\mathrm{WSe}_2$-encapsulated graphene \cite{christoph} at electron density $n_0\approx 5\times 10^{11}~{\rm cm}^{-2}$ and room temperature ($300~{\rm K}$). In the geometry in Fig.~\ref{fig:setup}, we choose $d=5~{\rm nm}$, $w=5~{\rm \mmu m}$, and $L=1~{\rm \mmu m}$.\footnote{The mass $m$ in this system is set by the Fermi velocity $v_{\mathrm{F}}$ and Fermi momentum $p_{\mathrm{F}}$ as $m=p_{\mathrm{F}}/v_{\mathrm{F}}$.} Hydrodynamic effects do persist in this device despite the additional screening of the Coulomb interactions caused by small $d$. (In fact, it has been experimentally checked that they persist even in extreme devices with $d$ down to $d \sim 1~{\rm nm}$ \cite{Manchester}.)  When $d=5$ nm, the mean free path for momentum conserving electron-electron collisions is $\ell_{\mathrm{ee}}\approx 500$ nm \cite{Manchester}, which is smaller than the device geometry;  therefore, we expect hydrodynamic effects to be of relevance \cite{mendl}.  (By increasing $d$, $\ell_{\mathrm{ee}}$ can be made even smaller).  In this setup, we estimate (see SI) that $v_{\mathrm{s}}\approx 1.4\times 10^6$ m/s, $\tilde \eta \approx 0.03$, and $\tilde\gamma \approx 0.11$.   The critical bias current beyond which instability occurs is $\tilde v_0 \approx 0.09$. Such bias current is accessible in experiment without undue heating of the phonons \cite{meric}.

\begin{figure*}
\centering
\subfloat[frequency]{\includegraphics[width=3.3in]{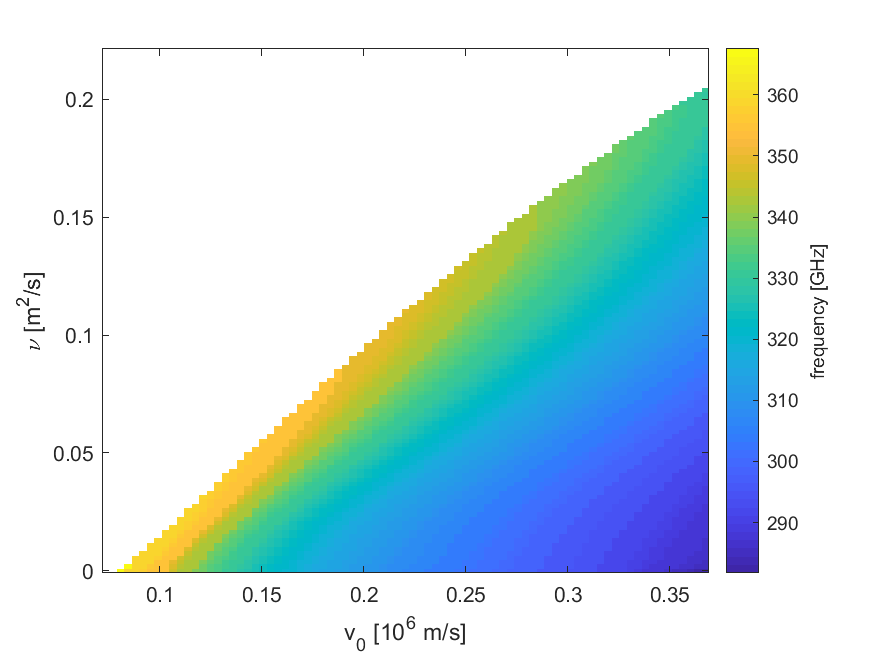}\label{fig:freq}}%
\hspace{0.05\textwidth}%
\subfloat[power]{\includegraphics[width=3.3in]{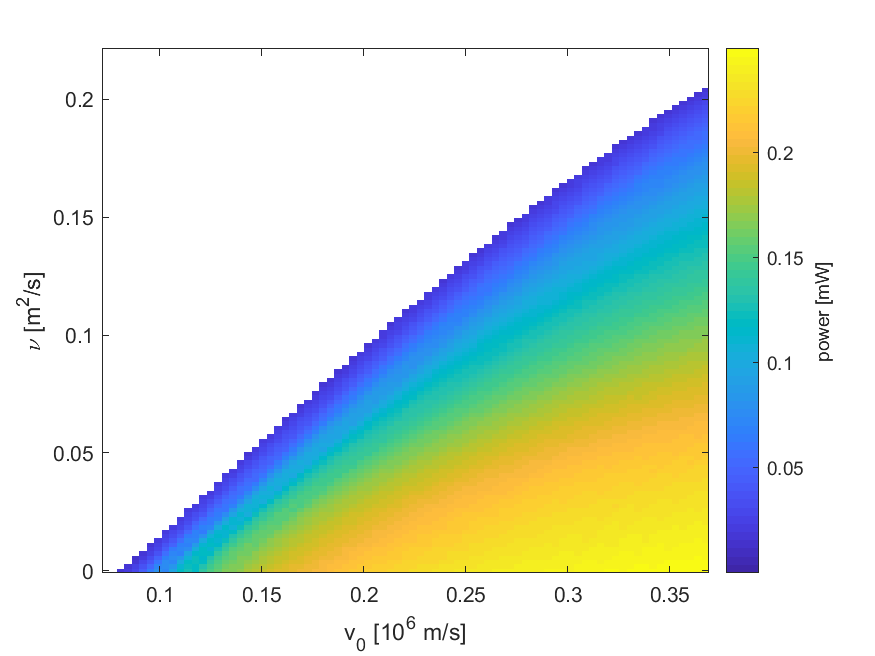}\label{fig:power}}
\caption{(a) The frequency of the emitted radiation as a function of $\nu$ and $v_0$; and (b) the maximal power dissipated, calculated from Eq.~\eqref{eq:Pupperbound}, as a function of $\nu$ and $v_0$. We have fixed $\gamma = 0.16~{\rm ps}^{-1}$ in both subplots. Parts of parameter space shaded white do not have an instability.}
\end{figure*}

Fig.~\ref{fig:freq} shows the frequency $f$ of the hydrodynamic oscillator as a function of $v_0$ and $\nu$, at the endpoint of the instability, defined by $f = 1/T$.  A key prediction is that as $v_0$ increases, the frequency of oscillations always decreases. Indeed, at the onset of the instability, $f=v_{\mathrm{s}}/(4L)$ and is set by the oscillation time of a quarter-wavelength fluctuation. In our proposed experiment, $f\approx 375~{\rm GHz}$ lies at the lower end of the THz gap \cite{Dhillon_jphysD_2017}.   Our simulations suggest that adding charge (increasing $n$) to the cavity takes longer at higher $v_0$; the time to remove this extra charge is very slightly decreasing at higher $v_0$, thus increasing the period and decreasing the frequency of the oscillator.  We propose that this increased charging time is a consequence of the additional formation time of the shock wave which moves to the left, but do not have an analytic understanding of this effect.

We also expect the qualitative sign change of this nonlinear shift in the frequency of radiation is robust and independent of the precise mechanism of coupling to an antenna. This is guaranteed if the feedback from dynamical electromagnetism onto the hydrodynamic equations of motion is a small effect. Moreover, in our proposed device the gate distance (5 nm) is so small relative to the electronic mean free path ($\ell_{\rm ee}\sim 500~{\rm nm}$) that the Coulomb kernel from the electron liquid to the gate is essentially local.

Next, we calculate the maximum power radiated from an antenna in the setup of Fig.~\ref{fig:setup}.  (The precise power radiated will depend on details of the device, so the estimate below is optimal.)  Since the charge in the capacitor is $Q(t) = -e\int_0^L \mathrm{d}x n(x,t)$, we upper bound the power generated at the fundamental frequency by
\begin{equation}
P \le \frac{1}{T}\left|\int\limits_t^{t+T} \frac{\mathrm{d}t}{T} \frac{Q^2(t)}{2C} \cos\left( \frac{2\pi t}{T}\right)\right|. \label{eq:Pupperbound}
\end{equation}
Here, $C=\epsilon_{zz} \varepsilon_{0}Lw/d$ is the geometrical capacitance of the setup, $\epsilon_{zz}$ is the relative out-of-plane permittivity of the dielectric between graphene and the gate, and $\varepsilon_{0}=8.854\times 10^{-12}~{\rm F}/{\rm m}$ the free-space permittivity.

Fig.~\ref{fig:power} plots the upper bound \eqref{eq:Pupperbound} on the radiated power for $\epsilon_{zz}\approx3.3$, i.e.~$C\approx 3\times 10^{-14}~{\rm F}$. Clearly, the radiated power vanishes at the onset of the instability, and increases as the bias current further increases. For the experiment suggested above, the scale of radiated power deep in the unstable regime is $\sim 0.1~{\rm mW}$. For more modest bias currents, this level of radiation is reduced, but even extremely close to the onset of instability the radiation can exceed $\sim 0.01~{\rm mW}$. These output powers are easily detectable by using present day THz detectors, such as bolometers \cite{simoens} or antenna-coupled graphene field-effect transistors \cite{Vicarelli}.

\begin{figure}
\centering
\includegraphics[width=3.3in]{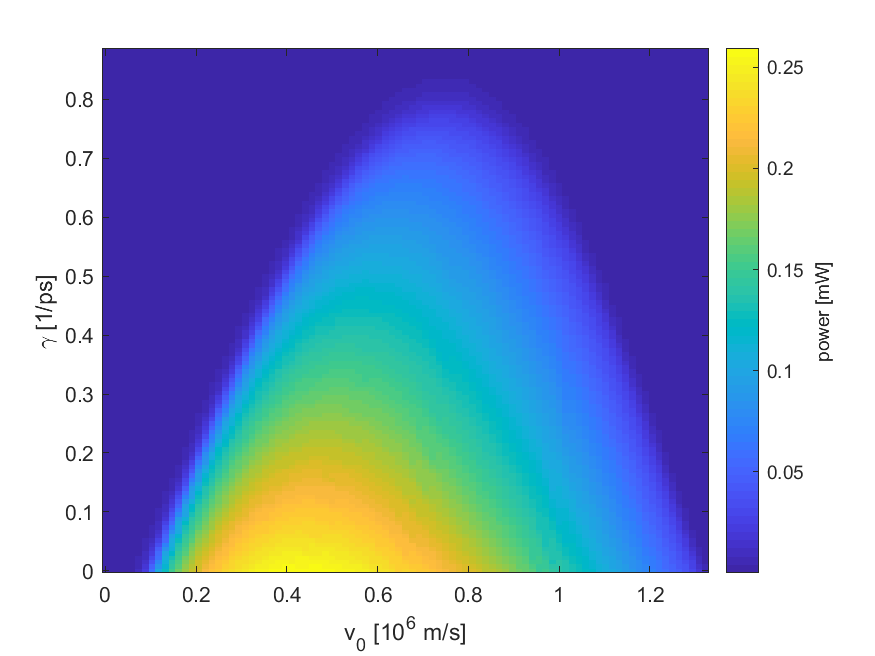}
\caption{For velocity $|v_0|<v_{\mathrm{s}}$, the instability becomes impossible to observe once momentum relaxation becomes too large.}
\label{fig:MRlimit}
\end{figure}

The most serious limitation for observing the instability in experiments is the presence of momentum relaxation. Fig.~\ref{fig:MRlimit} shows that beyond a critical value of $\gamma$, the instability disappears entirely. This can be understood crudely as follows: if $\gamma=\eta=0$, then $\mathrm{Im}(\omega_n) = \frac{v_{\mathrm{s}}}{2L} \log |\frac{v_{\mathrm{s}}+v_0}{v_{\mathrm{s}}-v_0}| < 0.45 \frac{v_{\mathrm{s}}}{L}$. Crudely estimating that momentum relaxation decreases $\mathrm{Im}(\omega_n)$ by $\frac{\gamma}{2}$ even when $v_{\mathrm{s}}\sim v_0$, we estimate that $\tilde \gamma < 0.9$ is required to see any instability. We numerically determined that $\tilde\gamma \lesssim 0.8$ was necessary to observe at least $0.03~{\rm mW}$ of radiated power. Keeping in mind that the hBN/$\mathrm{WSe}_2$-encapsulated graphene has a mean free path almost an order of magnitude larger than hBN-encapsulated graphene \cite{mayorov_nanolett_2012,wang_science_2013}, and that $\tilde\gamma \approx 0.1$ even in a relatively short device of $L=1$ $\mmu$m, it is quite possible that experiments on hBN-encapsulated graphene thus far could never observe the spontaneous instability.

A simple idea is to shrink $L$ to be small enough that $\gamma L \ll v_{\mathrm{s}}$. However, unless the electron-electron scattering rate is order(s) of magnitude faster than the electron-impurity scattering rate (which is not quite the case in present day materials), making $L$ very small will push the cavity into a completely ballistic regime: at least within linear response, electron-electron scattering is weak and the dynamics is well described by the kinetic theory of a fluctuating Fermi surface \cite{guo}. The instability can persist into this ballistic regime \cite{DSkin,mendl}, but its modeling requires a nonlinear solution of the Boltzmann equations for a Fermi liquid, beyond the scope of this Letter. The existence of instabilities in this kinetic regime may be very sensitive to boundary conditions \cite{mendl, satou}, which are difficult to control in experiment. Nevertheless, if the instability can persist deep into the ballistic limit with experimentally realized boundary conditions, then it may be preferable to work in this limit: in hBN/$\mathrm{WSe}_2$-encapsulated graphene, devices generating 1 THz radiation require $L\lesssim 300$ nm, which is not accurately described by hydrodynamics. Tuning the device length from $L=$100 nm to $L=1$ $\mmu$m generates radiation across the entire spectrum 0.3-3 THz.

\medskip

To conclude, in this paper we have described the endpoint of a hydrodynamic instability proposed by Dyakonov and Shur \cite{DS}, accounting for dissipative effects.  The frequency of the radiation generated at this endpoint is sensitive to microscopic parameters and bias current and can differ significantly from the linear response prediction.  Observing this nonlinear shift to the radiation frequency in experiment is a non-trivial consistency check with hydrodynamic theory beyond linear response and may open the door to future experiments in nonlinear hydrodynamics in solid-state devices.

\textit{Supplementary material.} The SI sketches a derivation of Eq.~\eqref{eq:LRomega}, provides details of the numerical method, shows simulation results for a more realistic pressure function for graphene, and investigates the effect of inhomogeneities in the coefficients of the hydrodynamic equations \eqref{eq:hydroeq}.

\textit{Data availability.} The data that support the findings of this study are available from the corresponding author upon reasonable request.

\textit{Acknowledgments.} We thank Kin Chung Fong, Christoph Stampfer, and Alessandro Tredicucci for useful discussions. C.M.\ would like to thank the Munich Quantum Center for support. A.L. was supported by the Gordon and Betty Moore Foundation's EPiQS Initiative through Grant GBMF4302. M.P. was supported by the European Union's Horizon 2020 research and innovation programme under grant agreement No.~785219 - GrapheneCore2.

\end{document}